\documentstyle[11pt,paspconf,epsf]{article}

\def \ltorder	{ \rlap{\lower .5ex \hbox{$\sim$} }{\raise .4ex \hbox{$<$} } }
\def \gtorder	{ \rlap{\lower .5ex \hbox{$\sim$} }{\raise .4ex \hbox{$>$} } }

\begin{document}

\title{The 24-Hour Night Shift:\\
Astronomy from Microlensing Monitoring Networks}

\author{Penny D. Sackett}
\affil{Kapteyn Astronomical Institute, 9700 AV Groningen, The Netherlands}

\begin{abstract}
Scores of on-going microlensing events are now announced yearly by the 
microlensing discovery teams OGLE, MACHO and EROS.  
These early warning systems have allowed other international 
microlensing networks to focus considerable resources on intense 
photometric --- and occasionally spectroscopic --- monitoring of 
microlensing events.  Early results include: 
metallicity measurements of main sequence Galactic bulge stars;     
limb darkening determinations for stars in the Bulge and 
Small Magellanic Cloud;  
proper motion measurements that constrain microlens identity; and  
constraints on Jovian-mass planets orbiting (presumably stellar) lenses. 
These results and auxiliary science such as variable star studies 
and optical identification of gamma ray bursts are reviewed.  
\end{abstract}



\section{Introduction}

Since the first Galactic events were announced  
(Alcock et~al.\ 1993; Aubourg et~al.\ 1993; Udalski et~al.\ 1993), 
microlensing has developed from an odd curiosity to a 
versatile astronomical tool.   Over the past 6 years, 
MACHO, OGLE and EROS have found hundreds of rare microlensing events 
in the dense stellar fields of the Galactic bulge and Magellanic 
Clouds --- a task truly akin to finding needles in the proverbial haystack.
Much of the progress that has been made in microlensing 
can be attributed to the ability and willingness of 
the discovery teams to provide public alerts 
of on-going microlensing events, allowing other resources to be dedicated 
to intense monitoring of large numbers of events.  
The additional astronomy derived from this often round-the-clock 
monitoring forms the subject of my brief review.   A more general 
discussion of microlensing can be found in the review by Mao (1999).

The scientific goals and capabilities of microlensing {\it discovery teams\/} 
and microlensing {\it monitoring teams\/} are quite different. 
The observational motivation of microlensing discovery projects is 
the search for very rare events lasting on the order of weeks to months, 
with the primary goal of understanding their relationship 
to the (dark and luminous) mass budget of the Galaxy.  
To this end, discovery teams 
(EROS II, MACHO, MOA, and OGLE II) must photometer huge numbers 
($\sim$10$^7$) of stars every 1$-$3 nights, using wide field 
($\sim$1 square deg) detectors on single, 
dedicated telescopes in Chile, Australia and New Zealand.

In contrast, the goal of intense monitoring networks 
is the detection and characterization of higher order perturbations --- 
or anomalies --- atop the primary microlensing signature.  Since these 
anomalies are often weak and of short duration, high photometric 
precision and dense temporal sampling are required.  To meet this challenge, 
international collaborations have built worldwide networks 
capable of precise, $\sim$hourly, round-the-clock monitoring.  
The current capabilities of these 
networks are summarized in Table~\ref{1999status}.  

\begin{table}
\caption{1999 Status of Microlensing Monitoring Teams\tablenotemark{a}} 
\label{1999status}
\begin{center}\footnotesize
\begin{tabular}{llccc}
Team & Telescope(s) & Time Allotment & Seeing & Pixel Size\\ 
\tableline
& & & & \\
GMAN\tablenotemark{b} & & & & \\
  & MSSSO 0.76m, Australia &~ 2 hrs/night 	& 2.3\arcsec & 0.36\arcsec    \\
  & CTIO 0.9m, Chile 	& 1.5 hrs/night         & 1.3\arcsec & 0.40\arcsec    \\
  & Wise 1.0, Israel 	& ~~1 obs/night         & 2.0\arcsec & 0.70\arcsec    \\
 & & & & \\
MOA\tablenotemark{c} & & & & \\
  & Mt. John 0.6m, New Zealand & 20 nights/year & 2.5\arcsec  & 0.81\arcsec    \\
 & & & & \cr
MPS\tablenotemark{d} &  & & & \\
  & MSSSO 1.9m, Australia & $\sim$8 weeks/year  & 2.3\arcsec    & --- \\
 & & & & \cr
PLANET\tablenotemark{e} &  & & & \\
  & SAAO 1.0m, South Africa& 100\% Bulge Season    & 1.6\arcsec & 0.31\arcsec \\
  & YALO 1.0m, Chile  	& ~75\% Bulge Season       & 1.5\arcsec & 0.30\arcsec \\
  & Canopus 1.0m, Tasmania & ~90\% Bulge Season    & 2.3\arcsec & 0.44\arcsec \\
  & Perth 0.6m, W. Australia & ~75\% Bulge Season  & 2.0\arcsec & 0.58\arcsec 
\end{tabular} 
\end{center}

\vskip -0.5cm
\tablenotetext{a}{Most teams expect improvements in capability and/or time allocations in 2000;\\ GMAN plans to stop normal operations at the 
end of the 1999 calendar year.}
\tablenotetext{b}{GMAN Alert Homepage:  http://darkstar.astro.washington.edu}
\tablenotetext{c}{MOA  Homepage:    http://www.phys.vuw.ac.nz/dept/projects/moa}
\tablenotetext{d}{MPS  Homepage:    http://bustard.phys.nd.edu/MPS}
\tablenotetext{e}{PLANET Homepage:  http://www.astro.rug.nl/$\sim$planet}
\end{table}

Due to the excellent real-time search and alert facilities of 
the discovery teams, intense monitoring networks can be productive 
continuously, typically following $\sim$10 events at any given time 
during the prolific Galactic bulge season. 
Because the location of the event is known,  
large detectors are not critical to the monitoring networks; more    
important is good image quality to ensure high  
photometric precision in dense microlensing fields.  
Due to the high temporal sampling rates required for  
anomaly detection, monitoring teams typically photometer 
${\cal O}(100)$ fewer stars ${\cal O}(20)$ times more frequently with 
${\cal O}(2)$ times better precision than do the discovery teams 
over the course of an observing season.   

\section{Science of Intense Microlensing Monitoring}

Intense microlensing monitoring science falls into three broad categories:  
(1) using microlensing as a large aperture, high-resolution telescope 
	to study the background sources,  
(2) characterizing anomalies in the light curve (or spectrum) 
	of the background source to learn about the lensing system, 
(3) auxiliary science from simultaneous monitoring or unrelated 
	time-critical photometry.	

\subsection{Learning about the Source: Abundances and Limb-Darkening}

Microlensing not only magnifies, but --- if the 
caustic structure\footnote{Caustics are sets of closed curves connecting 
positions in the source plane for which the determinant of the Jacobian 
lens mapping is zero.  Since the magnification is inversely 
proportional to the determinant, a source crossed by a caustic   
will experience exceedingly high magnification, which is finite thanks 
only to the finite size of the source itself.} 
from a multiple lens passes over the background star --- 
actually {\it differentially\/} 
magnifies and thus spatially resolves its stellar source.  
This makes microlensing an excellent high-magnification, high-resolution 
(though impossible to point!) telescope with which to study faint 
and distant sources.  

\begin{figure}
\plotfiddle{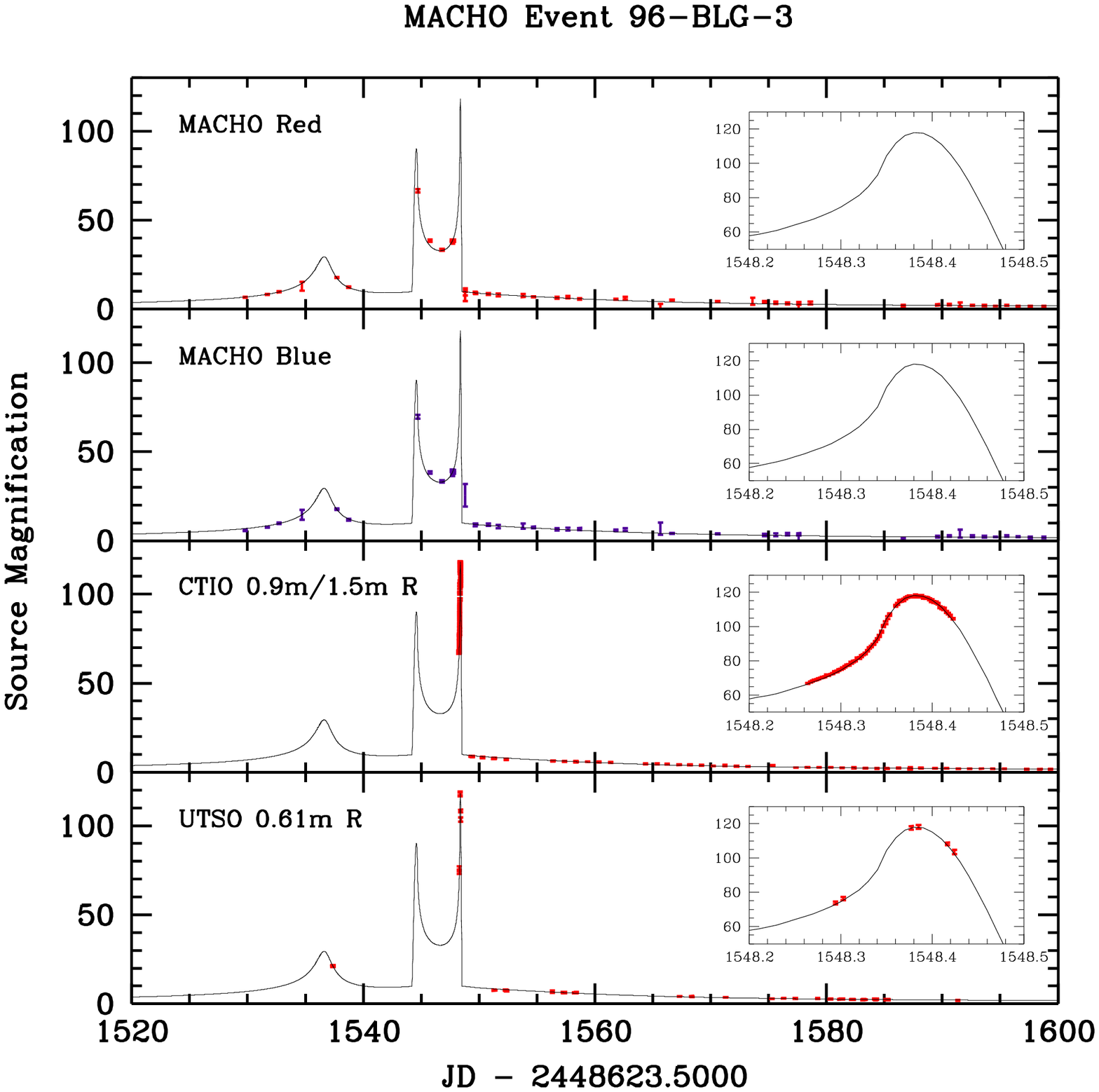}{8.6cm}{0}{36}{36}{-195}{-10}
\plotfiddle{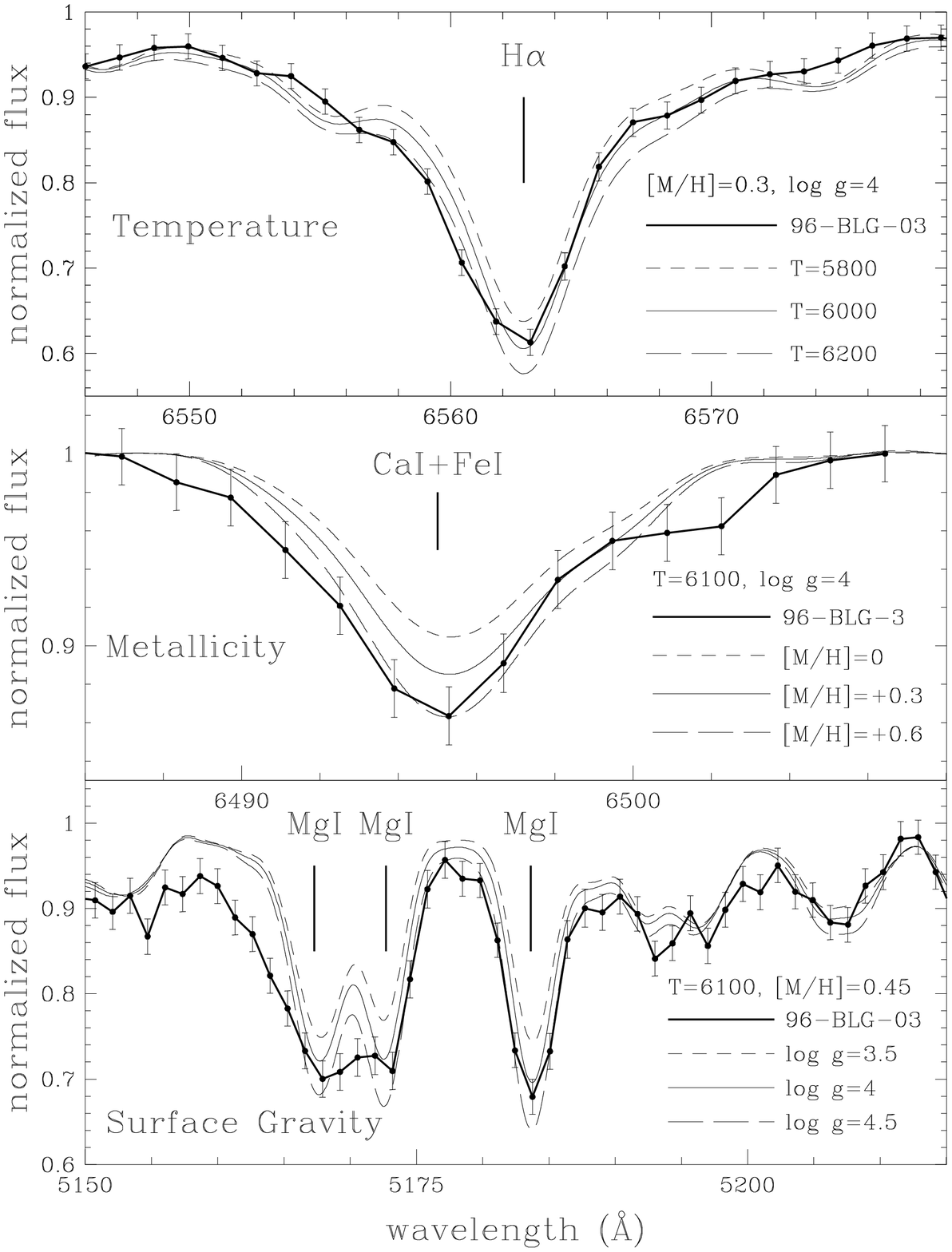}{8.6cm}{0}{30}{30}{15}{285}
\vskip -10.8cm
\caption{The high signal-to-noise target-of-opportunity 
NTT spectra of a bulge G-dwarf (right, Lennon et~al.\ 1996) 
were made possible by multi-site photometric monitoring 
by GMAN and an accurate prediction for the time 
of the caustic crossing (left, Alcock et~al.\ 2000).} \label{MB9603}
\end{figure}

Microlens telescopes have been used to acquire spectra of 
subgiant and dwarf stars in the Galactic bulge that are 
too faint and crowded to be studied in other way, 
thereby allowing age and metallicity determinations 
for some of the oldest stars in the center of our Galaxy 
(Lennon et~al., 1997).  
Real-time alerts of impending caustic crossings can be a boon to 
such work.  For example, the GMAN caustic alert for MACHO~96-BLG-3 
(Alcock et~al., 2000) allowed Lennon et~al.\ (1996) to 
measure the effective temperature, gravity and metallicity 
of the G-dwarf bulge source star; with the caustic boost, the 
NTT momentarily had the collecting power of a 17.5m telescope 
(Fig.~\ref{MB9603}).  A relatively modest factor of three 
microlensing boost aided Minitti et~al.\ (1998) to use the Keck to 
detect lithium in the turn-off source star of a different bulge event.

When a background star is transited by a caustic, 
the cool limb of the star will be magnified differentially 
during ingress and egress.  
This effect can be used to resolve the surface brightness profile of the 
source, yielding a limb-darkening measurement that can be 
checked against predictions from stellar atmospheric models.    
Except for the Sun and a few nearby super giants, such measurements 
are difficult to obtain any other way.

\begin{figure}
\vglue -1.3cm
\hglue 0.75cm\epsfxsize=0.53\textwidth\epsffile{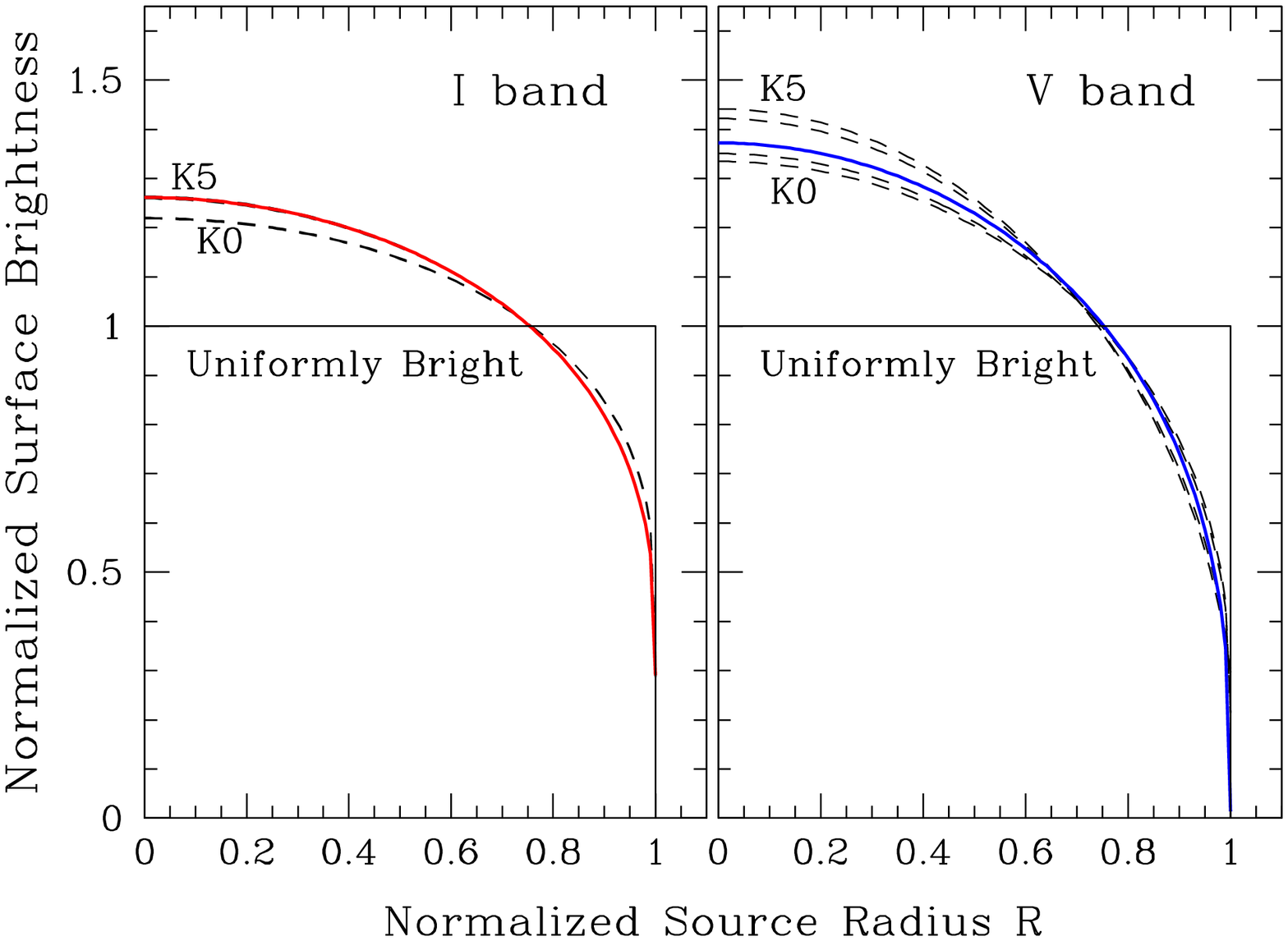}
\hglue -0.3cm\epsfxsize=0.53\textwidth\epsffile{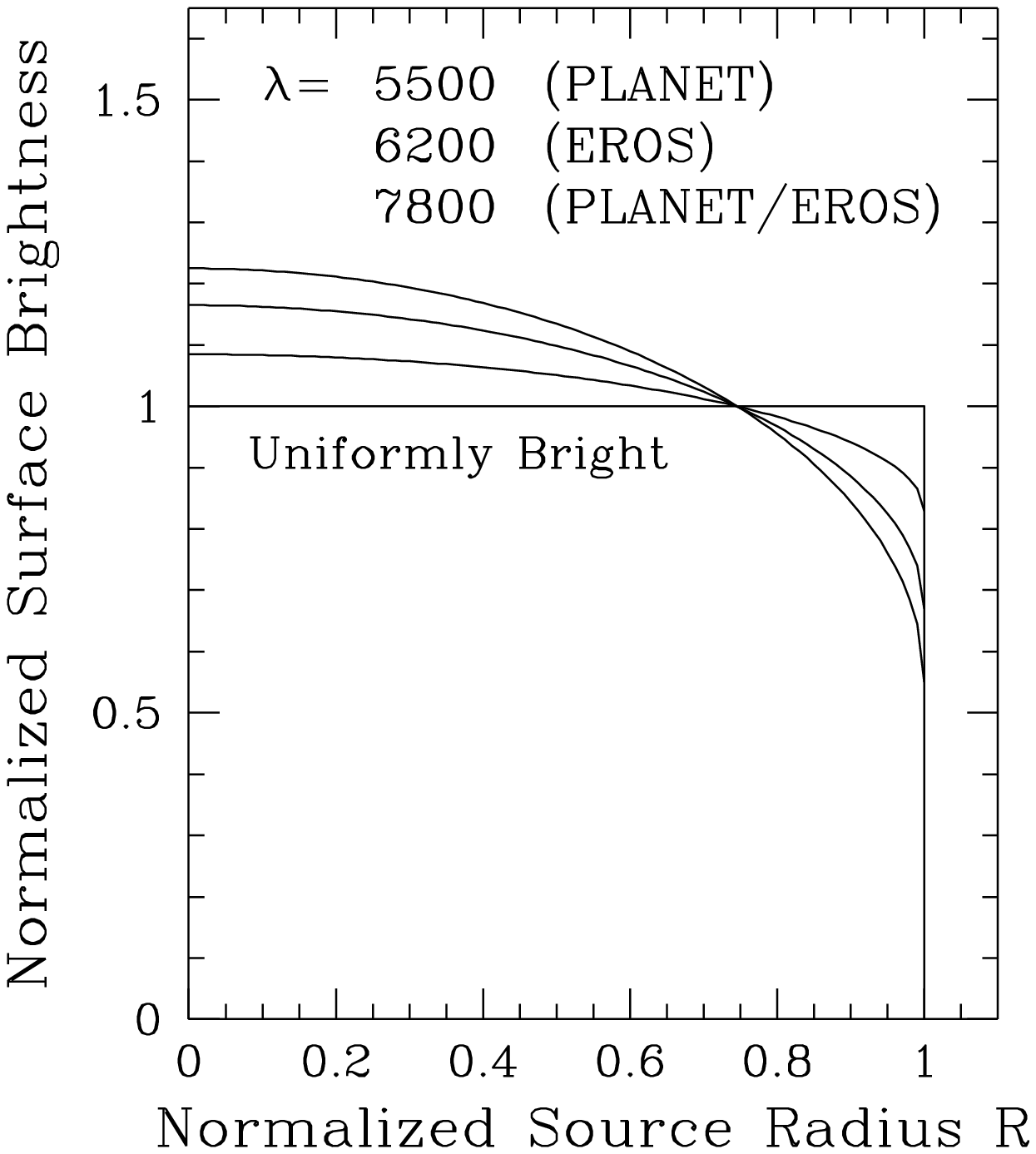}
\vglue -0.5cm
\caption{
{\it Left:\/}  The limb-darkened profiles reconstructed from 
PLANET light curve data (Albrow et~al.\ 1999b) 
for MACHO 97-BLG-28 (solid lines) 
are consistent with stellar atmosphere models (dashed lines) 
for the K2 giant source star. 
{\it Right:\/} Reconstructed stellar profiles in three passbands 
(lowest curve at star's center is reddest) 
for the metal-poor A-dwarf source of MACHO 98-SMC-1 
(Afonso et~al.\ 2000).   
}\label{limbdark}
\end{figure}

Round-the-clock monitoring of binary event MACHO~97-BLG-28 allowed 
the PLANET team to fulfill this scientific potential  
for the first time.  Data spaced by 3-30 minutes over 
the caustic peak allowed the slope change during the transit to be 
determined accurately, from which I- and V-band limb-darkening 
coefficients were derived directly without knowledge of 
the absolute stellar size (Fig.~\ref{limbdark}).  
The inferred surface brightness profile for the 
K-giant source star was consistent with stellar 
atmosphere models in both bands (Albrow et~al.\ 1999b).  
An analysis of the binary microlensing event MACHO~98-SMC-1 using 
data from five microlensing teams has produced limb-darkening 
coefficients for an A-dwarf source star 
in three passbands (Afonso et~al.\ 2000).  This SMC star has an angular 
radius of only {\it $\sim$80 nanoarcseconds\/}.  
No appropriate atmospheric models could be found for comparison. 
As expected, the A-dwarf is less limb-darkened than the K-giant, and 
limb darkening is less pronounced in redder passbands (Fig.~\ref{limbdark}).   
Limb-darkening measurements can be expected at the rate of $\gtorder$2  
per year; PLANET has at least two suitable light curves from its 1999 season.

These first studies show that microlensing can be used effectively 
to boost the signal and resolution of our earth-bound 
telescopes.  Since caustic structure is observed in $\sim$10\% of all microlensing events, and the approximate timing of a caustic crossing is 
often known a few days in advance, the primary hurdle will be  
re-assignment of telescope time on short-time scales for follow-up studies.
Facilities that are flexible enough will be 
rewarded not only with more discoveries like these but also --- 
as many of the contributions at this meeting have shown --- 
the possibility to study stellar spots, rotation, and polarization, 
or even resolve quasar sizes and 
the atmospheric line structure of background source stars.

\subsection{Learing about the Lens: Kinematics and Companions}

If the nature of the background source is sufficiently well-understood  
(or irrelevant), a light curve generated by a caustic  
passing near the source can be used to learn  
about the lens responsible for the caustic structure.   
In good circumstances, three quantities can be determined: 
$\rho_*$, the size of the source star in units of the angular 
Einstein ring radius $\theta_{\rm E}$  
($\theta_{\rm E} \equiv \sqrt{4 G M\, (D_S - D_L)/ (c^2  D_S\, D_L)}$ 
for a lens of mass $M$ a distance $D_L$ from the observer and 
$D_S - D_L$ from the source); 
$b$, the instantaneous separation of a static binary lens in 
units of $\theta_{\rm E}$; and $q$, the mass ratio of the binary.  
If either the Earth or the binary lens executes a significant fraction 
of its orbit during the duration of the microlensing event, 
gentle light curve anomalies may yield additional kinematic information 
(eg., Alcock et~al.\ 1995).  However, by using the stellar radius as a 
micro ruler and a caustic transit as a micro clock, the relative 
proper motion $\mu_{\rm LS}$ between source and lens can sometimes be  
determined even in short duration events.

\begin{figure}
\vglue -2.6cm
\hglue 0.55cm\epsfxsize=0.52\textwidth\epsffile{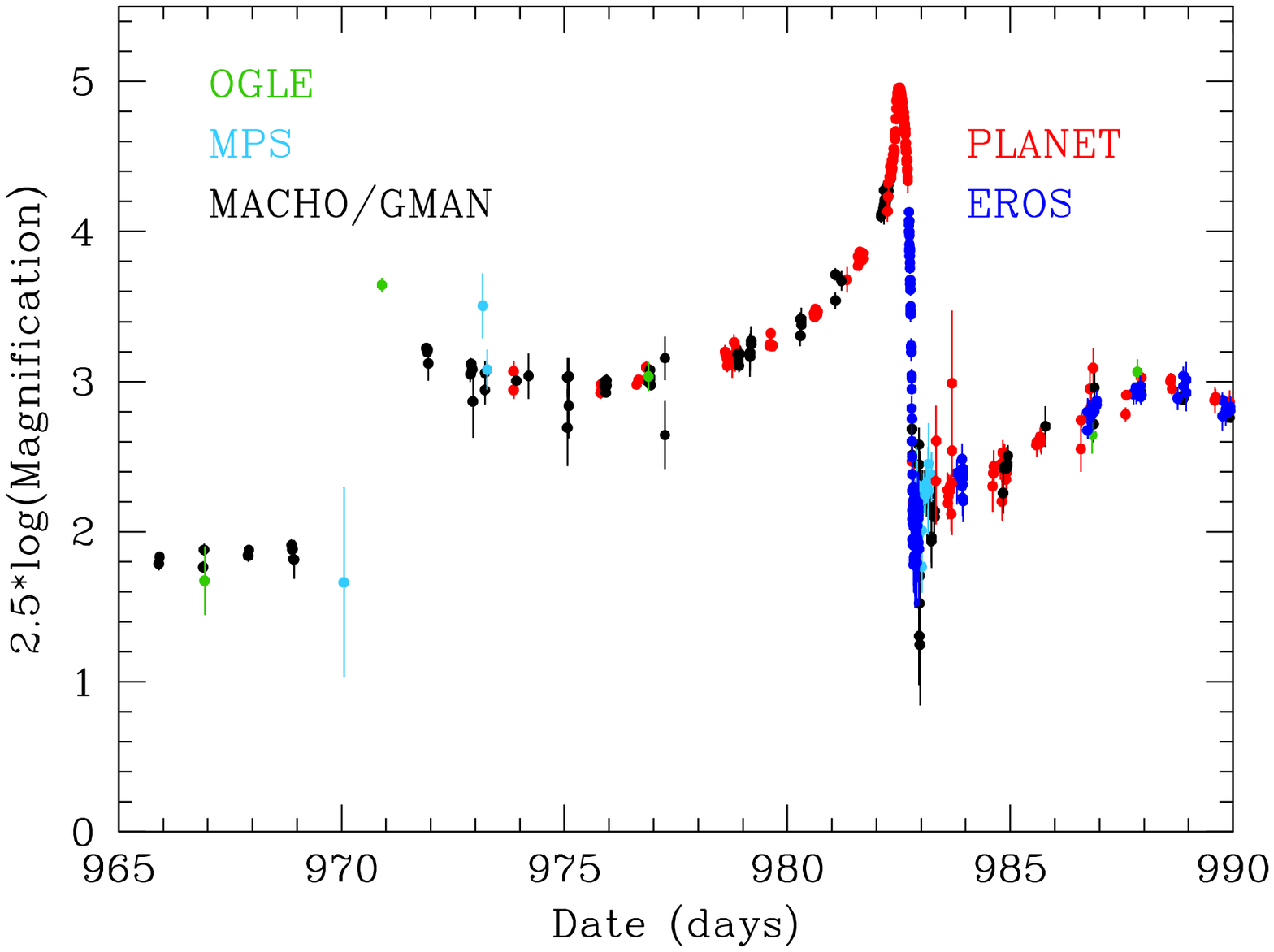}
\hglue -0.2cm\epsfxsize=0.48\textwidth\epsffile{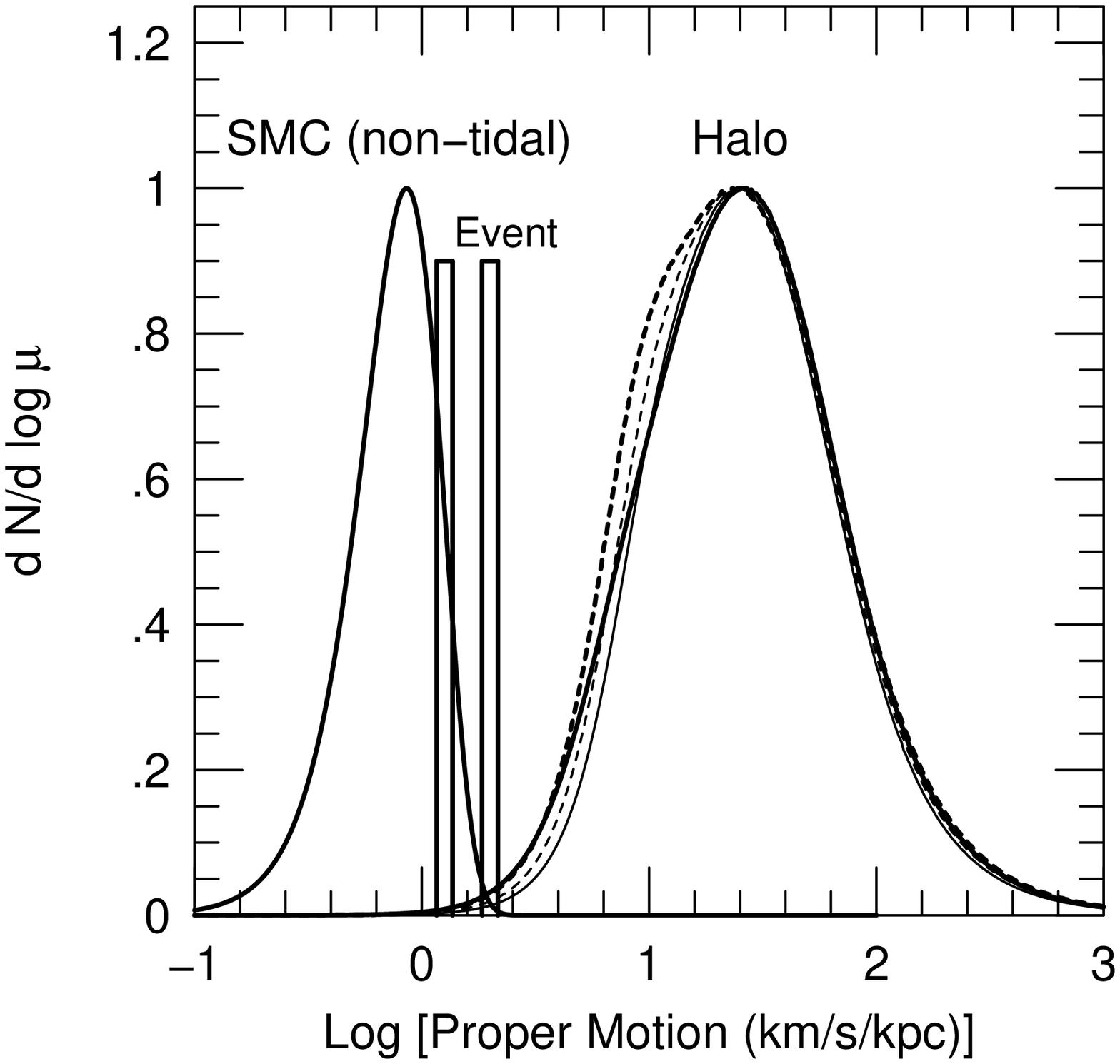}
\vglue -0.5cm
\caption{The light curve of MACHO~98-SMC-01 (left) 
combines data from five microlensing teams.  Dense sampling 
by PLANET and EROS over the second peak produced the limb-darkening 
measurements of Fig.~\ref{limbdark}.  Data from MACHO/GMAN, MPS 
and OGLE near the first peak ruled out the 
higher $\mu_{\rm LS}$ model of PLANET (right, Albrow et~al.\ 1999a), 
securing the result that its relative proper motion 
places the lens in the SMC, not the Galactic dark halo.
}\label{propermotion}
\end{figure}
 
This possibility prompted five different teams 
(Afonso et~al.\ 1998, Udalski et~al.\ 1998, 
Albrow et~al.\ 1999a, Alcock et~al.\ 1999, Rhie et~al.\ 1999) 
to monitor the rare binary microlensing event MACHO~98-SMC-01 
in the direction of the Small Magellanic Cloud (SMC).  
The time taken by the SMC source star to cross  
the line caustic could be measured directly from 
the caustic peak portion of the light curve. The rest of the 
light curve yielded the angle between source trajectory and 
caustic, allowing computation of the time $t_*$ taken by the source to  
traverse a distance equal to its own radius.  When combined with 
the angular stellar diameter $\theta_*$ estimated from spectral 
typing or color-surface brightness relations, the relative 
lens-source proper motion $\mu_{\rm LS} \equiv \theta_*/t_*$ was  
determined.  A joint analysis (Afonso et~al.\ 2000) has confirmed the 
independent conclusions of all five groups that the lens proper motion 
is too small to have the halo-like kinematics expected 
for Galactic dark matter (Fig.~\ref{propermotion}), but is quite 
reasonable for a normal stellar binary lens residing in the SMC itself.
More binary events in the direction of the Clouds must be studied 
in order to determine whether this conclusion is representative 
of the whole class, which would cast doubt on the notion 
that microlenses constitute a major fraction of the Galactic 
dark matter. 
     
\begin{figure}
\hglue 2.0cm\epsfxsize=0.70\textwidth\epsffile{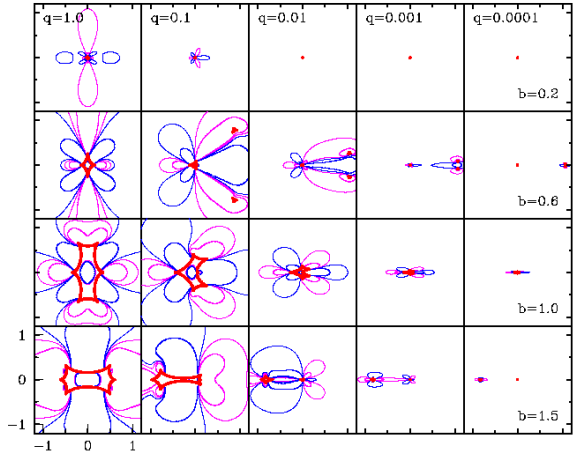}
\vglue -3.3cm
\caption{Contours of excess magnification ($\pm$1\%, $\pm$5\%, bold $= 
\infty = $caustics) for binary lenses of different mass ratio $q$ and 
separation $b$. Light curves of background point sources 
with trajectories crossing these contours will deviate from that of a 
single lens by the indicated amount.  Distances are in $\theta_{\rm E}$. 
Adapted from Gaudi \& Sackett (2000).
}\label{binarypanel}
\end{figure}
 
Although some binary light curves do not allow a unique determination of 
the angular separation $b$ and mass ratio $q$, 
in most well-sampled cases categorization into stellar (q\gtorder 0.1) 
or planetary (q\ltorder 0.01) companions can be made easily.  
(Note that the mass of Jupiter $M_{\rm J} \approx 0.001\, M_{\odot}$.) 
As Fig.~\ref{binarypanel} 
shows, for stellar lensing binaries of moderate separations 
(0.5 \ltorder b \ltorder 1.5), nearly all source trajectories 
inside $\theta_{\rm E}$ will produce light curve anomalies at 
the $> 1$\% level.  Many stellar binaries have been detected in 
this way.  In other microlensing events, 
the absence of detectable anomalies places strong constraints 
on the presence of lensing companions in this range of $q$ and $b$.  

\begin{figure}
\vglue -0.1cm
\hglue -0.1cm\epsfxsize=0.51\textwidth\epsffile{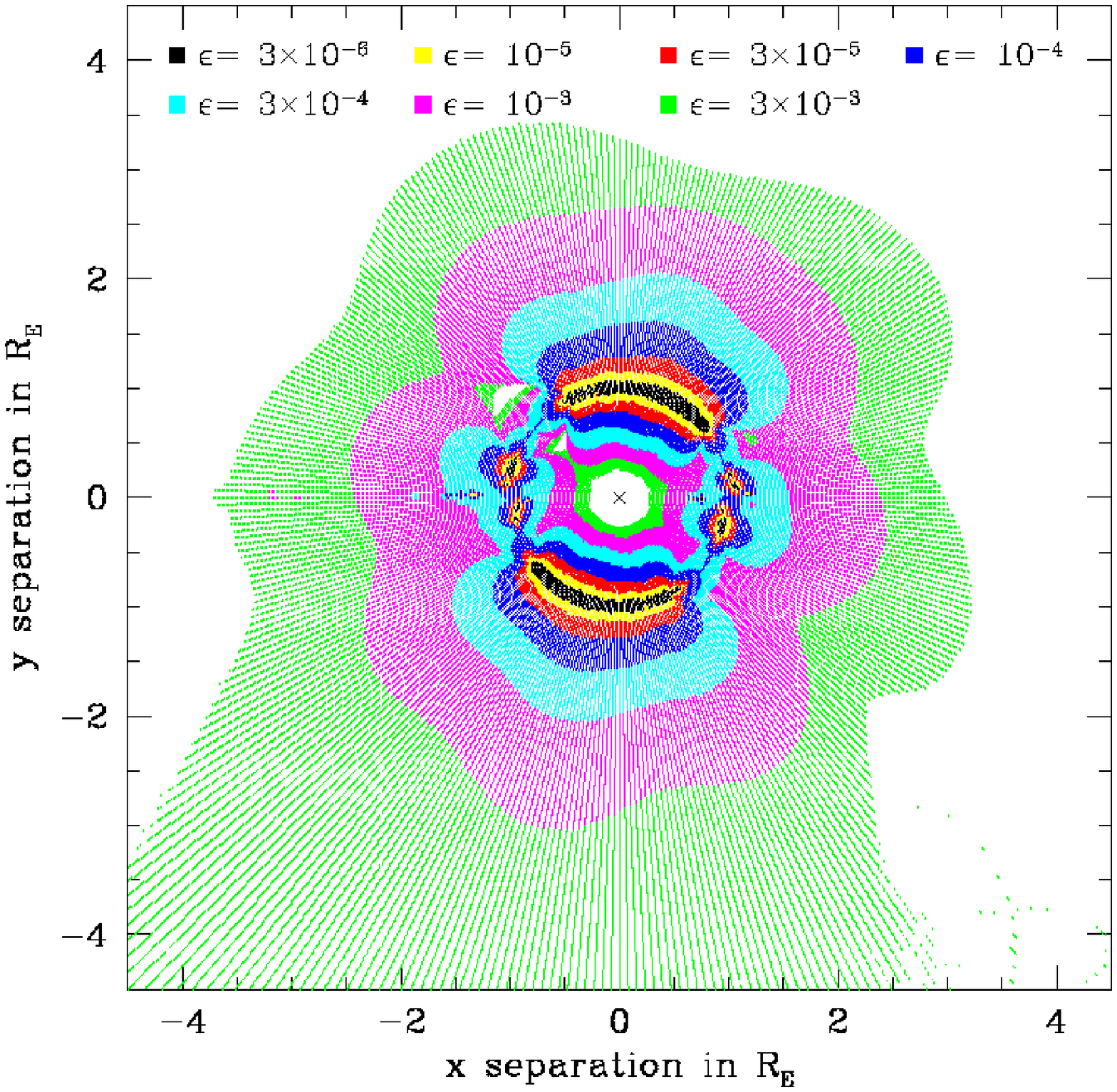}
\vglue -7.0cm
\hglue 6.5cm\epsfxsize=0.53\textwidth\epsffile{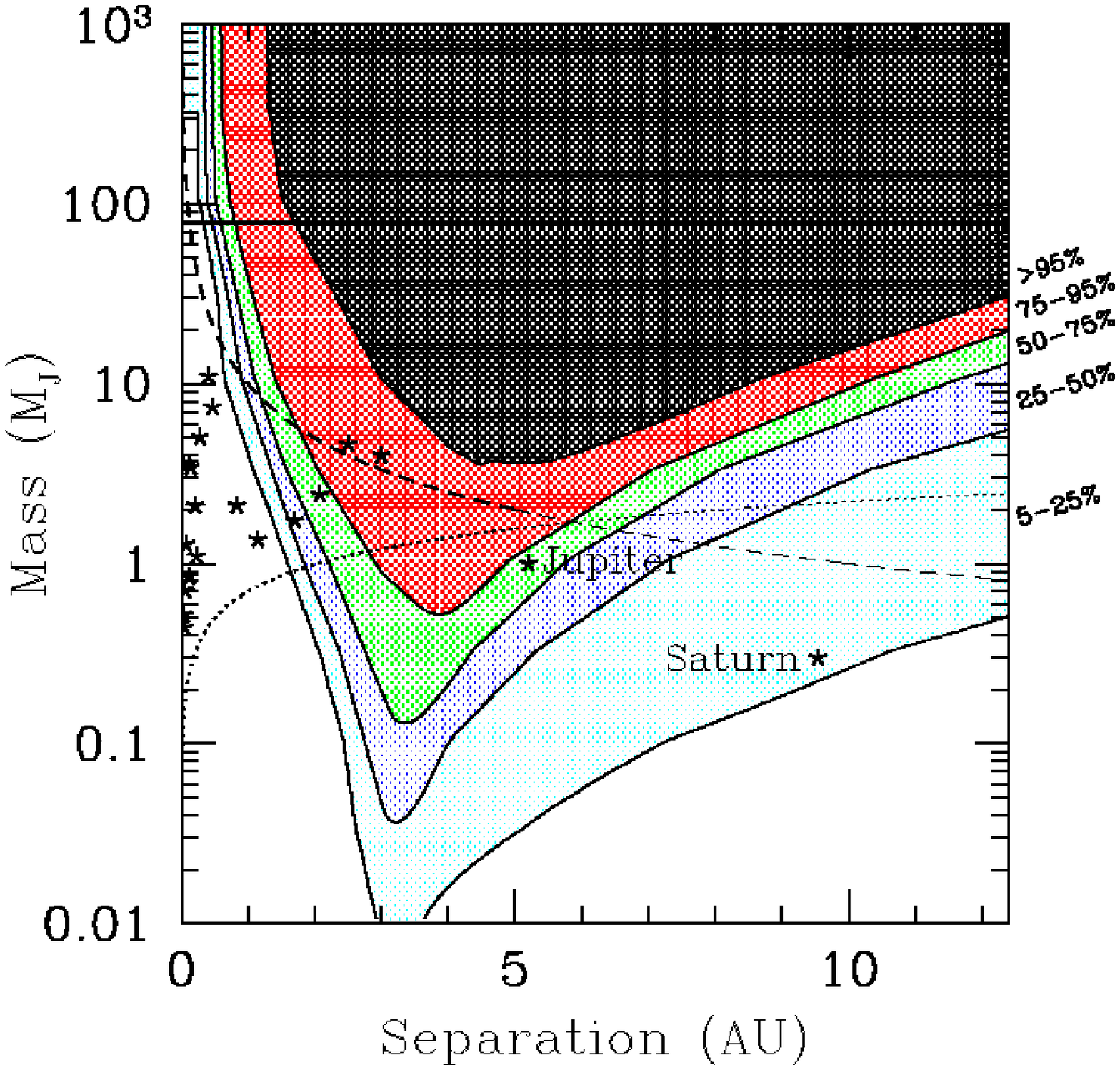}
\vglue -0.25cm
\caption{Exclusion diagrams for companions to two different microlenses. 
{\it Left:\/} Companions to MACHO 98-BLG-35  
are excluded in the shaded regions (Rhie et al.\ 2000); the 
larger the mass ratio, the larger the excluded area on the sky. 
{\it Right:\/} Exclusion contours at various significance levels for  
OGLE-1998-BUL-14 companions of given mass and orbital radii, 
for a lens with $M = 1 M_{\odot}$ and 
$D_L \, \theta_{\rm E} = 3.1$AU (Albrow et al.\ 2000a).  
Asterisks denote known radial velocity planets.
}\label{nojupiters}
\end{figure}

To date, no clear planetary microlensing signatures have been detected.  
Generally, planets are difficult to rule out (or detect) because 
the regions of significant anomalous magnification are small.
Nevertheless, a large class of massive planets have been 
ruled out in two high-magnification events that exhibited no large deviations.  
The MOA and MPS monitoring teams have excluded the presence 
of massive companions with $q > 0.001$ and $0.4 < b < 2.5$ 
in MACHO-98-BLG-35 (Rhie et al.\ 2000).  In this very 
high magnification event, companions with $q = 0.003$ were ruled  
out to $\sim$3 Einstein radii (Fig.~\ref{nojupiters}).  
The 600 data points collected by the PLANET team for OGLE-98-BUl-14 
excluded companions with $q > 0.01$ over $0.4 < b < 2.4$ 
(Albrow et al.\ 2000a).   

Assuming reasonable microlens characteristics, the exclusion 
(and detection) regime of microlensing extends to higher orbital 
separation than that of the radial velocity method (right panel, 
Fig.~\ref{nojupiters}). 
Microlensing is also capable of detecting planets of smaller mass 
ratio, although care must be taken to remove systematic effects such as  
variable seeing (Albrow et al.\ 2000a), since perturbations caused by 
low-mass ($q < 0.001$) planets typically will be mild and short-lived. 
Microlensing is thus an excellent complement 
to Doppler searches for extra-solar planets in the Galaxy.  
A very preliminary statistical analysis of several light curves 
from the 1998 PLANET data set indicates that $\sim$25\% of 
all lenses do not have companions with mass $\gtorder$5~$M_{\rm J}$ 
orbiting within an annulus bounded by $\sim$1 -- 6~AU 
(Gaudi et al. 1999, this meeting).  For orbital radii close to the 
Einstein ring ($\sim$2.5~AU), this excluded fraction rises to $\sim$50\%.  
Considerably tighter constraints on the presence of Jovian planets 
orbiting stellar microlenses are expected when analysis of the 
full PLANET data set is complete. 
 
\subsection{Auxiliary and Serendipitous Science: Variables and GRB}

The frequent, high-precision photometry of dense fields required  
of intense microlensing monitoring is ideally suited for the study 
of rare, faint, short-period, or low-amplitude variable stars.  
The PLANET collaboration has discovered Bulge 
variables as faint as I = 19, as rapidly-varying as once per 2 hours, 
and as subtle as 0.04~mag in total amplitude  
(Albrow et al.\ 2000b).  The global, rapid-response nature of the  
monitoring networks can be used for target-of-opportunity studies 
of rare, irregular variables such as fading gamma ray bursts (GRBs).  
PLANET photometry produced precise optical 
positions for GRB~990510 (Vreeswijk et al.\ 1999) and GRB~990712 
(Bakos et al.\ 1999), enabling VLT spectroscopy hours later that 
yielded redshifts for both bursts.

\section{The Best is Yet to Come}

Due to its limited scope, this review has centered only on the 
observational rewards reaped by the global networks 
that stand watch 24 hours a day monitoring microlensing events 
for light curve anomalies.  It is the theorists, however, who have 
led the way, with ideas often thought to be far-fetched when 
proposed, yet verified a few short years later.  
If recent history is any guide, boldness of vision combined with care 
in execution will continue to serve theorist and observer alike 
in this young and rapidly evolving field of astrophysics.
 
\vskip 0.75cm

\acknowledgments

It is a pleasure to thank my colleagues in the PLANET collaboration 
for permission to discuss some of our work before publication, and 
Andy Becker and Phil Yock for providing detailed information and figures  
related to the microlensing monitoring efforts of which they are part. 
I am grateful to the organizing committee, Kapteyn Institute, and 
the LKBF for travel support.


\begin{references}
\reference Afonso et~al.\ (EROS) 1998, A\&A, 337, L17 		 
\reference Afonso, C., et~al.\ (EROS, MACHO/GMAN, MPS, OGLE \& PLANET) 2000, 
  ApJ, in press (astro-ph/9907247) 
\reference Albrow, M.D., et~al.\ (PLANET) 1999a, ApJ, 512, 672 	 
\reference Albrow, M.D., et~al.\ (PLANET) 1999b, ApJ, 522, 1011  
\reference Albrow, M.D., et~al.\ (PLANET) 2000a, ApJ, in press 
  (astro-ph/9909325) 
\reference Albrow, M.D., et~al.\ (PLANET) 2000b, 
	Impact of Large Scale Surveys on Pulsating Star Research, 
	L. Szabados \& D. Kurtz, San~Francisco: ASP 
\reference Alcock, C., et~al.\ (MACHO) 1993, Nature, 365, 621
\reference Alcock, C., et~al.\ (MACHO) 1995, ApJ, 454, L125 
\reference Alcock, C., et~al.\ (MACHO/GMAN) 1999, ApJ, 518, 44 
\reference Alcock, C., et~al.\ (MACHO/GMAN) 2000, ApJ, in press 
  (astro-ph/9907369)
\reference Aubourg, E., et~al.\ (EROS) 1993, Nature, 365, 623
\reference Bakos, G., et~al.\ 1999, GCN 387			
\reference Gaudi, B.S. \& Sackett, P.D. 2000, ApJ, 529, 000 (astro-ph/9904339)
\reference Lennon, D.J., et~al.\ 1996, ApJ, 471, L23
\reference Lennon, D.J., et~al.\ 1997, Messenger, 90, 30 (astro-ph/9711147)
\reference Mao, S.\ 1999, these proceedings (astro-ph/9909302)
\reference Minitti, D. et~al.\ 1998, ApJ, 499, L175
\reference Rhie S., et~al.\ 1999 (MPS), ApJ, 522, 1037		
\reference Rhie S., et~al.\ 2000 (MOA \& MPS), ApJ, in press 
    (astro-ph/9905151)       
\reference Udalski, A., et~al.\ (OGLE) 1993, Acta Astron., 43, 289
\reference Udalski A., et~al.\ (OGLE) 1998, AcA, 48, 431	
\reference Vreeswijk, P., et~al.\ 1999, GCN 310			
\end{references}
\end{document}